\documentclass[a4paper]{article}

\usepackage[english]{babel}
\usepackage[utf8x]{inputenc}
\usepackage[T1]{fontenc}
\usepackage{verbatim}

\usepackage[a4paper,top=3cm,bottom=2cm,left=3cm,right=3cm,marginparwidth=1.75cm]{geometry}

\usepackage{amsmath}
\usepackage{amsfonts} %for mathbb
\usepackage[numbered]{mcode}
\usepackage{graphicx}
\usepackage[colorinlistoftodos]{todonotes}
\usepackage[colorlinks=true, allcolors=blue]{hyperref}
\usepackage{subfloat}
\usepackage{subfigure}

\title{Open 2D Electrical Impedance Tomography data archive}
\author{Andreas Hauptmann, Ville Kolehmainen,  Nguyet Minh Mach,\\Tuomo Savolainen, Aku Seppänen, Samuli Siltanen}

\newcommand{\Addresses}{{% additional braces for segregating \footnotesize
  \bigskip
  \footnotesize

  Andreas Hauptmann, \textsc{Department of Mathematics and Statistics, University of Helsinki, Finland}\par\nopagebreak
  \textit{E-mail address}:  \texttt{andreas.hauptmann@helsinki.fi}

  \medskip
  
 Ville Kolehmainen, \textsc{Department of Applied Physics, University of Eastern Finland}\par\nopagebreak
  \textit{E-mail address}:  \texttt{ville.kolehmainen@uef.fi}

  \medskip
  Nguyet Minh Mach, \textsc{Department of Mathematics and Statistics, University of Helsinki, Finland}\par\nopagebreak
  \textit{E-mail address}: \texttt{minh.mach@helsinki.fi}

\medskip

  Tuomo Savolainen, \textsc{Department of Applied Physics, University of Eastern Finland}\par\nopagebreak
  \textit{E-mail address}: \texttt{tuomo.savolainen@uef.fi}
  
  \medskip

  Aku Seppänen, \textsc{Department of Applied Physics, University of Eastern Finland}\par\nopagebreak
  \textit{E-mail address}: \texttt{aku.seppanen@uef.fi}
  
  \medskip

  Samuli Siltanen, \textsc{Department of Mathematics and Statistics, University of Helsinki, Finland}\par\nopagebreak
  \textit{E-mail address}: \texttt{samuli.siltanen@helsinki.fi}
}}

\begin{document}
\maketitle

\begin{abstract}
This document reports an Open 2D Electrical Impedance Tomography (EIT) data set.
The EIT measurements were collected from a circular body (a flat tank filled with saline) with various choices of conductive and resistive inclusions. Data are available at \url{http://fips.fi/EIT_dataset.php} and can be freely used for scientific purposes with appropriate references to them, and  to this document at \url{https://arxiv.org}. The data set consists of (1) current patterns and voltage measurements of a circular tank containing different targets, (2) photos of the tank and targets and (3) a MATLAB-code for reading the data. 
%This is the documentation of the Electrical Impedance Tomography data set of a circular reference body and different choices of targets in 2D. Data are available at \url{http://fips.fi/EIT_dataset.php} and can be freely used for scientific purposes with appropriate references to them, and  to this document at \url{https://arxiv.org}. The data set consists of (1) current patterns and voltage measurements, (2) photos of the tank and targets and (3) a suggested MATLAB-code to read the data. 
A video report of the data collection session is available at \url{https://www.youtube.com/watch?v=65Zca_qd1Y8}.
\end{abstract}

%%%%%%%%%%%%%%%%%%%%%%
%%%%%%%%%%%%%%%%%%%%%%

\section{Introduction}

Electrical Impedance Tomography (EIT) is a diffusive imaging modality in which the spatial
distribution of electric conductivity (or its reciprocal, resistivity) is reconstructed from
a set of current injections and potential measurements from electrodes attached on its
surface.

\begin{figure}[htp]
	\begin{center}
		\subfigure[Experiment 1.0; homogeneous]{\label{fig:2a}\includegraphics[width=7cm]{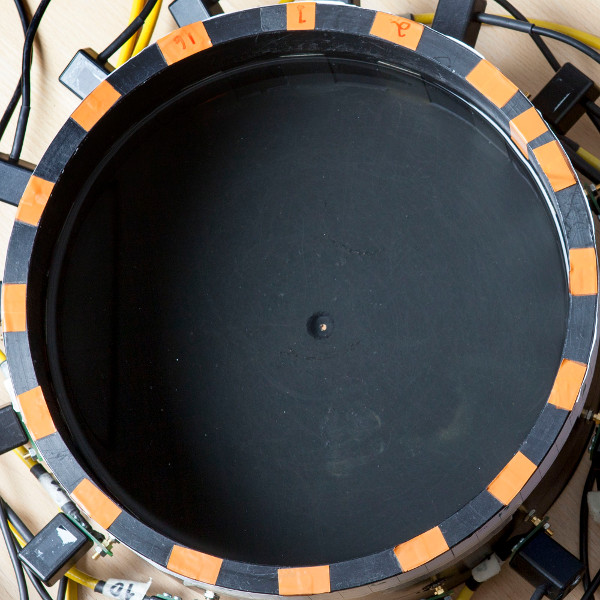}}
		\subfigure[Experiment 2.1; two plastic inclusions]{\label{fig:2b}\includegraphics[width=7cm]{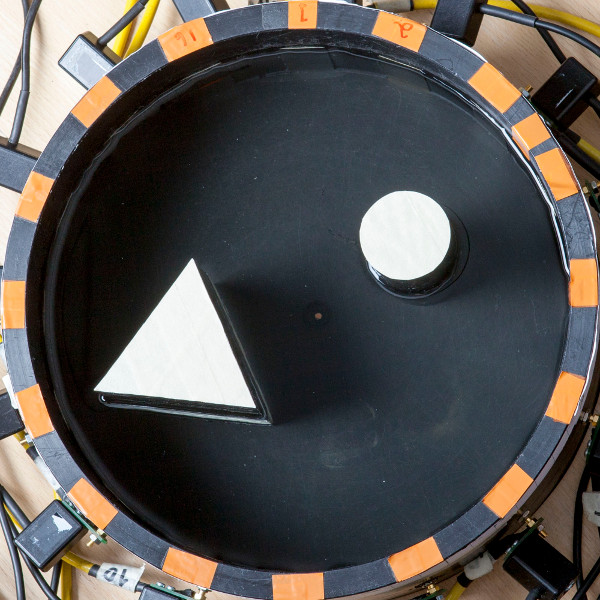}}
	\end{center}
\mbox{}\vspace{-5mm}
	\caption{Experimental setup for collecting 2D EIT data.}
	\label{fig:2}
\end{figure}

The purpose of this project is to disseminate EIT measurement data from simple experimental setups for the use of testing and development of image reconstruction methods.
The experiments were carried out using a flat cylindrical tank filled with saline.
Electrically conductive and resistive inclusions were inserted inside the tank,
so that they extended from the bottom of the tank to the saline top surface
(For two examples, see Figure \ref{fig:2}.).
As also the electrodes on the inner surface of the tank extended from top to bottom,
the geometry was essentially two-dimensional (2D) and
the measurement data is suitable for 2D EIT.

%The main idea behind the project was to create 2D phantom experimental data for testing inclusions reconstruction algorithms in electrical impedance tomography. Targets with different shape, size, conductivity, position were put in a circular tank filled with salt water. Small alternating currents with low  frequencies were injected through electrodes attached on the surface of the tank and voltage measurement were collected at the electrodes. 

The rest of this report is organized as follows.
In Section \ref{Sec:measurements},
we briefly describe the EIT measurements and the target objects used in the experiments.
In Section \ref{sec.Data}, we provide instructions for downloading the Open EIT data,
and briefly describe the contents of the data files.
Finally, in Section \ref{sec.ExampleReconstructions},
we show example reconstructions computed based on the data.

%%%%%%%%%%%%%%%%%%%%%%
%%%%%%%%%%%%%%%%%%%%%%

\section{Test cases and data collection} 
%EIT measurements}
\label{Sec:measurements}

\begin{figure}[htp]
\centering
\includegraphics[angle=0,origin=c,width=0.8\textwidth]{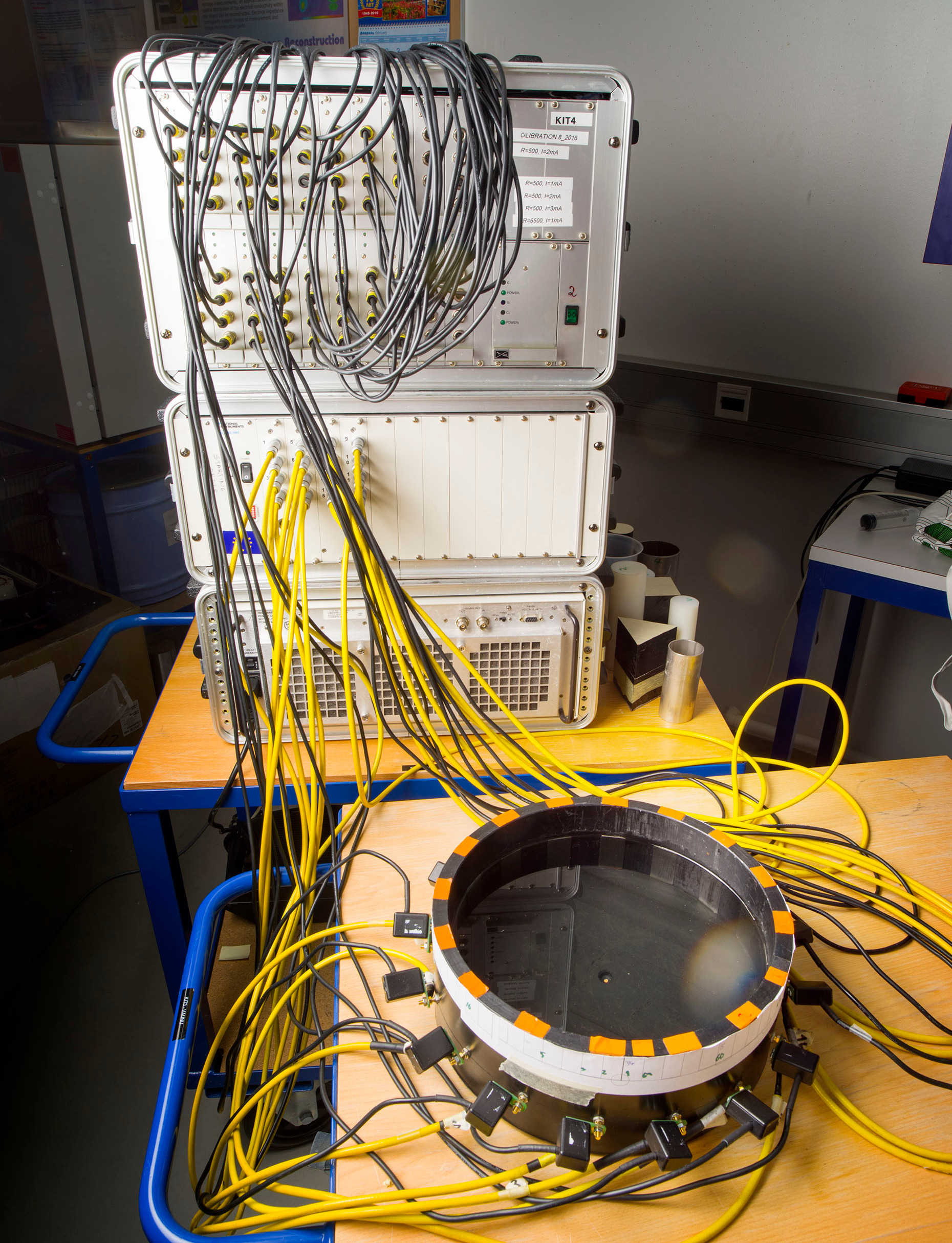}
\caption{The KIT4 measurement system at the University of Eastern Finland}\label{fig:KIT4}
\end{figure}

The EIT data set was measured using the KIT4 (Kuopio Impedance Tomography) measurement system (Figure \ref{fig:KIT4}).  The system consists of three parts. The top part shown in the figure is the voltage measurement module with $80$ channels. Up to $80$ voltage signals can be simultaneously transferred from the electrodes to the module through the black cables.  The middle part is  the current injection module with $16$ independent current injection channels connected with $16$ electrodes via the yellow cables. The lowest part is the controller module.  The reader is referred to \cite{kourunen2008suitability} for detailed description of the KIT4 system.

The experiments were conducted using a tank of circular cylinder shape. The diameter of the tank was $28$ cm. 
Sixteen rectangular electrodes (height $7$ cm, width $2.5$ cm) made of stainless steel were attached equidistantly on the inner surface of the tank.
The electrodes were marked with orange tapes on the tank wall,
and they were numbered in clockwise order, see Figure \ref{fig:2}.
In this and in the remaining figures of this document, 
the topmost electrode is labeled as Electrode $1$.

\begin{figure}[!]
    \centering
    \begin{minipage}{.25\textwidth}
        \centering
        \includegraphics[width=3cm]{photos/fantom_1_0.jpg} 
{\bf Case 1.0:}  Homogeneous target
\vspace{5mm}
    \end{minipage}
    \begin{minipage}{0.25\textwidth}
        \centering
        \includegraphics[width=3cm]{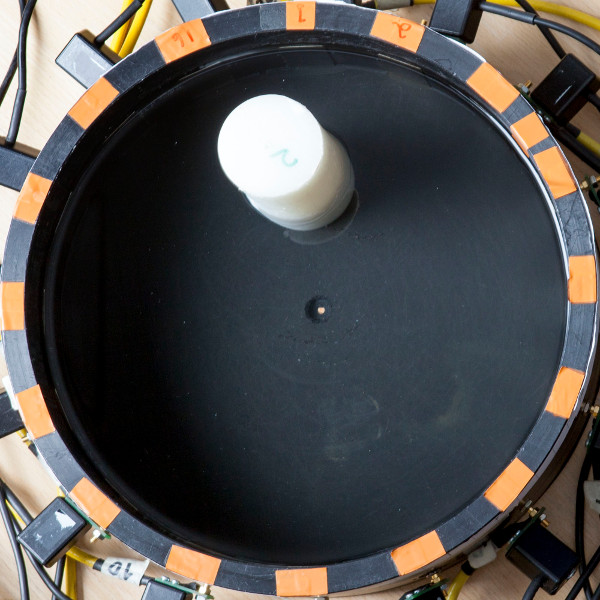} 
  {\bf Case 1.1:} 1 plastic inclusion (circular)
\vspace{5mm}
    \end{minipage}
    \begin{minipage}{0.25\textwidth}
        \centering
        \includegraphics[width=3cm]{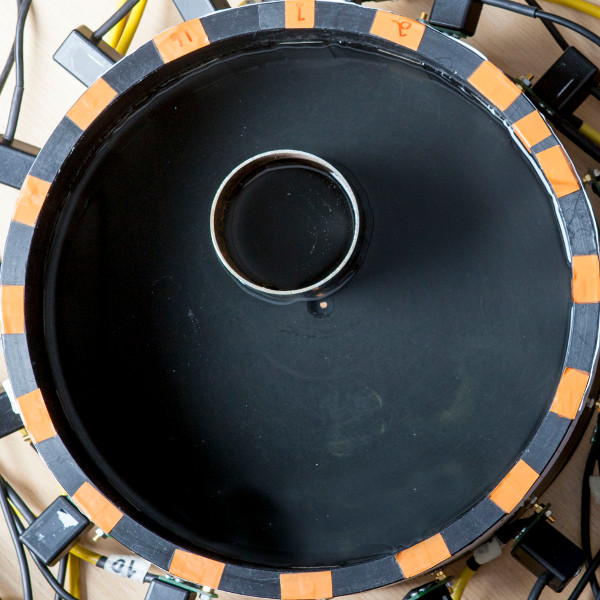} 
  {\bf Case 1.2:} 1 metallic inclusion (circular, hollow)
    \end{minipage}
    \begin{minipage}{0.25\textwidth}
        \centering
        \includegraphics[width=3cm]{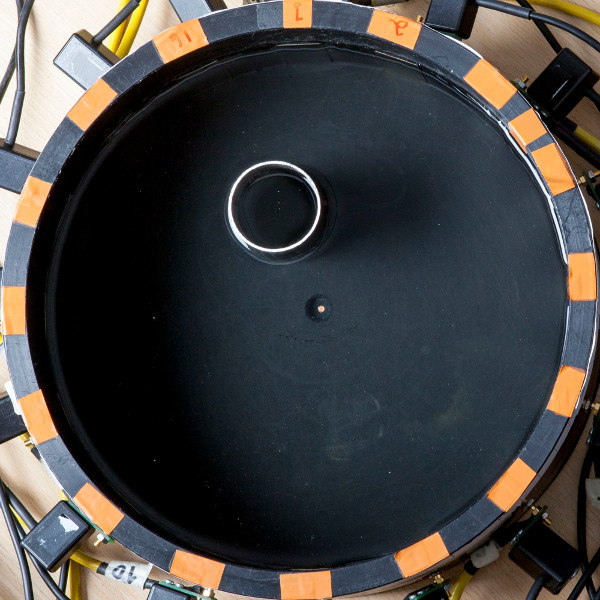} 
  {\bf Case 1.3:} 1 metallic inclusion (circular, hollow)
    \end{minipage}
    \begin{minipage}{0.25\textwidth}
        \centering
        \includegraphics[width=3cm]{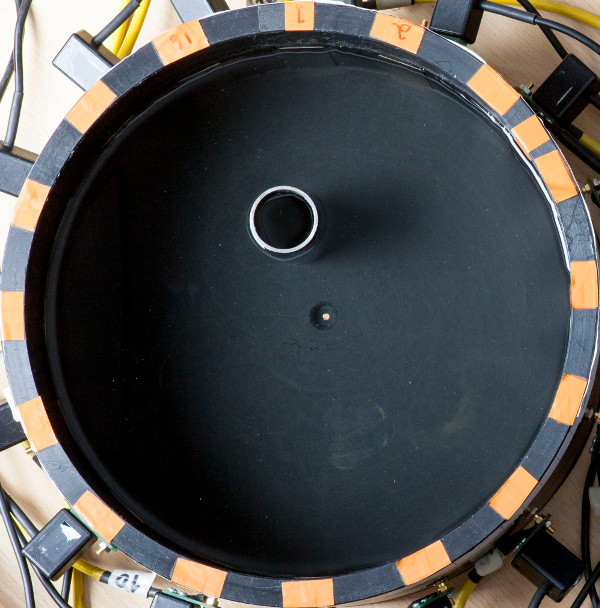} 
  {\bf Case 1.4:} 1 metallic inclusion (circular, hollow)
    \end{minipage}
\caption{Photographs and descriptions of the targets, Cases 1.0 -- 1.4.
\label{fig.photos1}
}
\end{figure}

\begin{figure}[!]
    \centering
    \begin{minipage}{0.25\textwidth}
        \centering
        \includegraphics[width=3cm]{photos/fantom_2_1.jpg} 
  {\bf Case 2.1:} 2 plastic inclusions (triangular \& circular)
\vspace{5mm}
    \end{minipage}
    \begin{minipage}{0.25\textwidth}
        \centering
        \includegraphics[width=3cm]{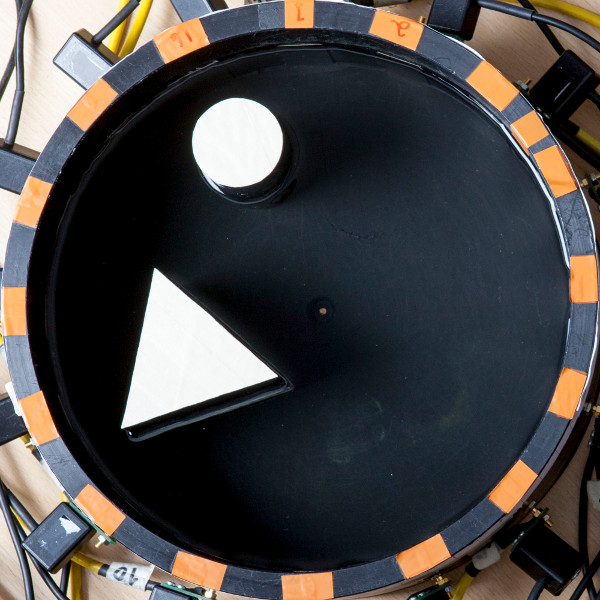} 
  {\bf Case 2.2:} 2 plastic inclusions (triangular \& circular)
\vspace{5mm}
    \end{minipage}
    \begin{minipage}{0.25\textwidth}
        \centering
        \includegraphics[width=3cm]{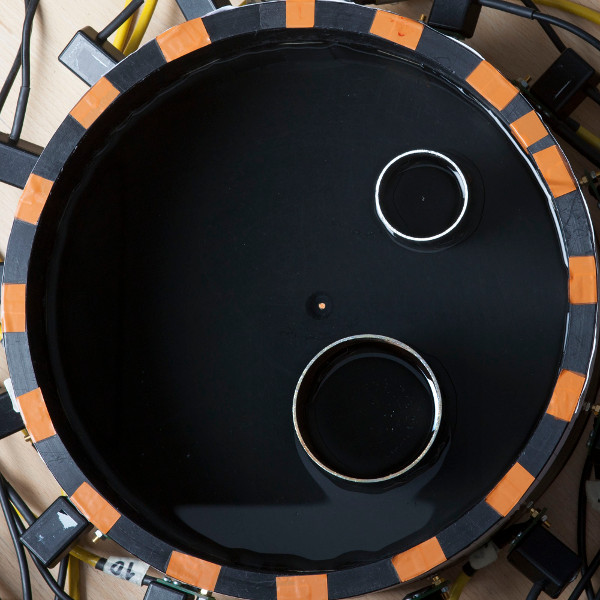} 
  {\bf Case 2.3:} 2 metallic inclusions (circular, hollow)
\vspace{5mm}
    \end{minipage}
    \begin{minipage}{0.25\textwidth}
        \centering
        \includegraphics[width=3cm]{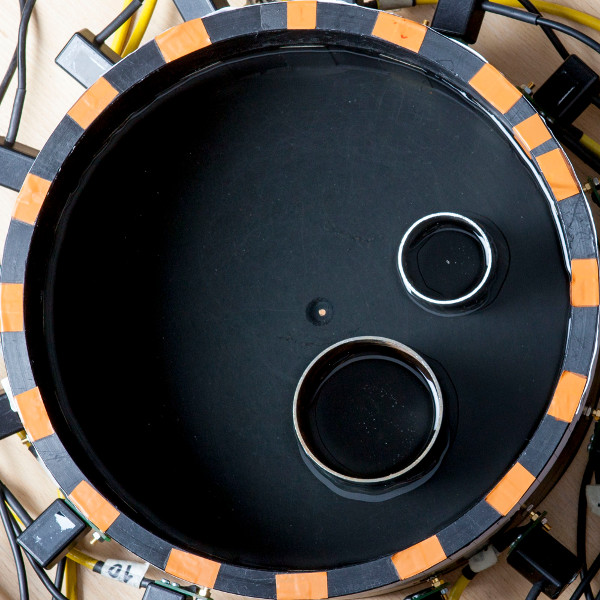} 
  {\bf Case 2.4:} 2 metallic inclusions (circular, hollow)
    \end{minipage}
    \begin{minipage}{0.25\textwidth}
        \centering
        \includegraphics[width=3cm]{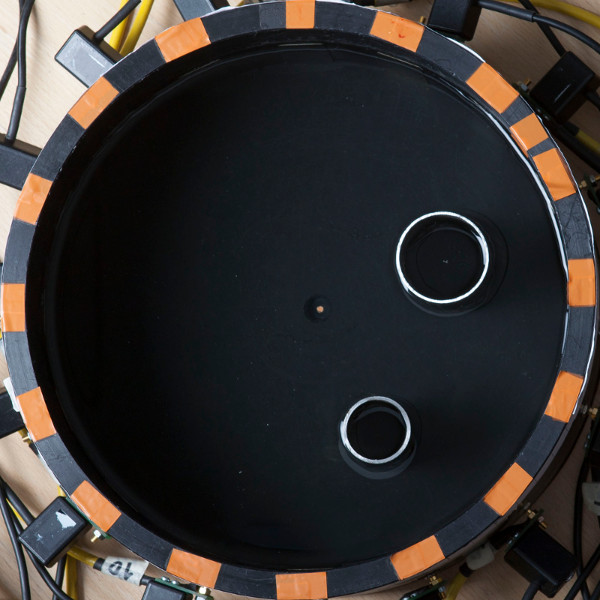} 
  {\bf Case 2.5:} 2 metallic inclusions (circular, hollow)
    \end{minipage}
    \begin{minipage}{0.25\textwidth}
        \centering
        \includegraphics[width=3cm]{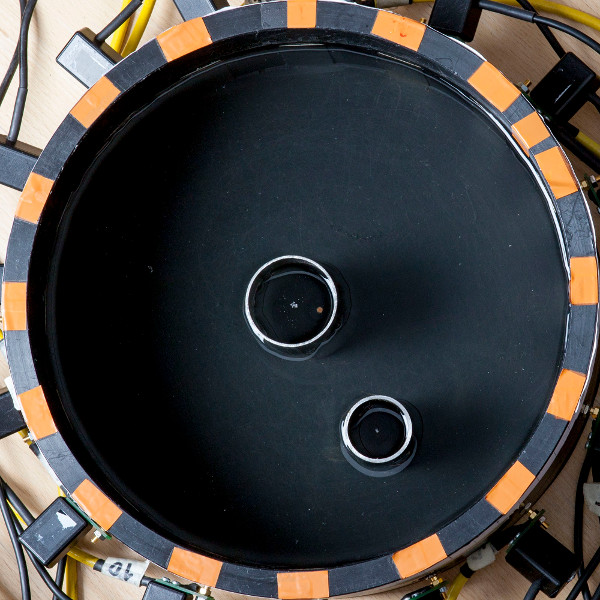} 
  {\bf Case 2.6:} 2 metallic inclusions (circular, hollow)
    \end{minipage}
\caption{Photographs and descriptions of the targets, Cases 2.1 -- 2.6.
\label{fig.photos2}
}
\end{figure}

\begin{figure}[!]
    \centering
    \begin{minipage}{0.25\textwidth}
        \centering
        \includegraphics[width=3cm]{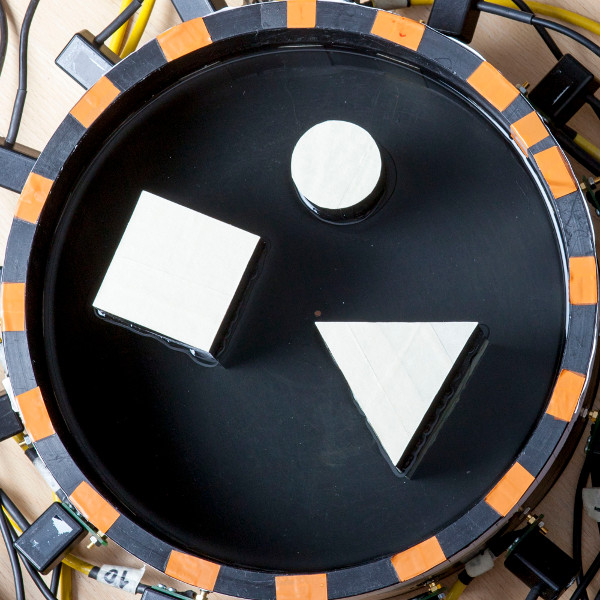} 
  {\bf Case 3.1:} 3 plastic inclusions (rectangular, triangular \& circular)
\vspace{5mm}
    \end{minipage}
    \begin{minipage}{0.25\textwidth}
        \centering
        \includegraphics[width=3cm]{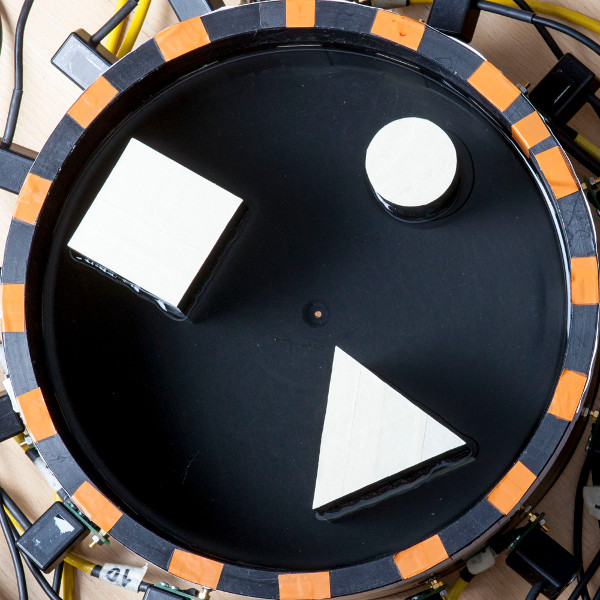} 
  {\bf Case 3.2:} 3 plastic inclusions (rectangular, triangular \& circular)
\vspace{5mm}
    \end{minipage}
    \begin{minipage}{0.25\textwidth}
        \centering
        \includegraphics[width=3cm]{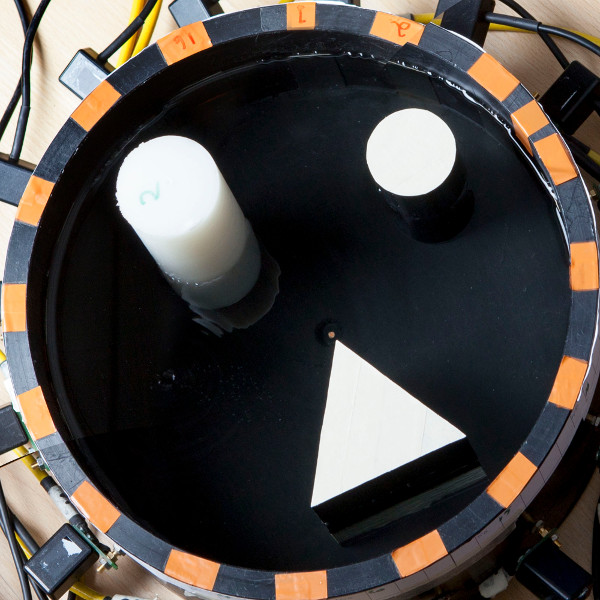} 
  {\bf Case 3.3:} 3 plastic inclusions (triangular \& 2 circular)
\vspace{5mm}
    \end{minipage}
    \begin{minipage}{0.25\textwidth}
        \centering
        \includegraphics[width=3cm]{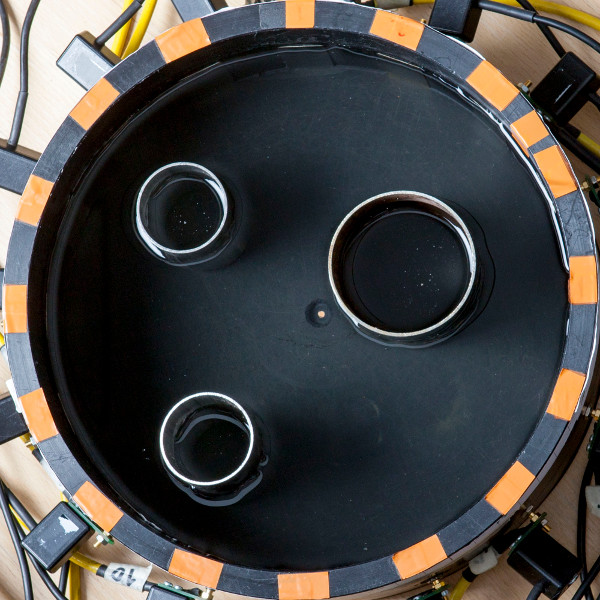} 
  {\bf Case 3.4:} 3 metallic inclusions (circular, hollow)
    \end{minipage}
    \begin{minipage}{0.25\textwidth}
        \centering
        \includegraphics[width=3cm]{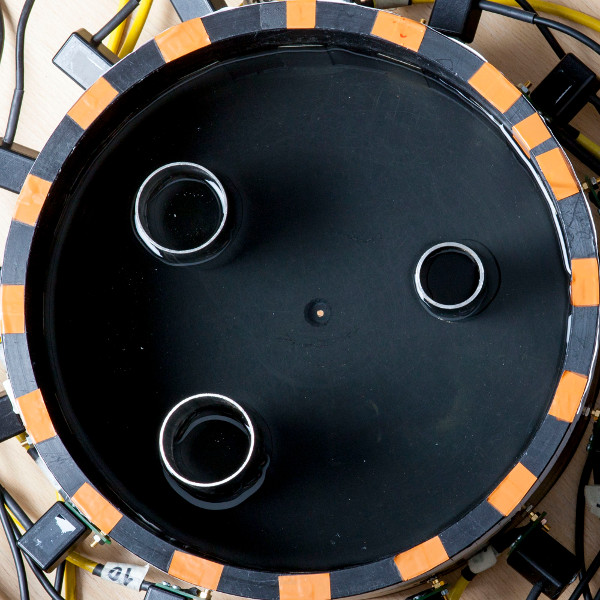} 
  {\bf Case 3.5:} 3 metallic inclusions (circular, hollow)
    \end{minipage}
    \begin{minipage}{0.25\textwidth}
        \centering
        \includegraphics[width=3cm]{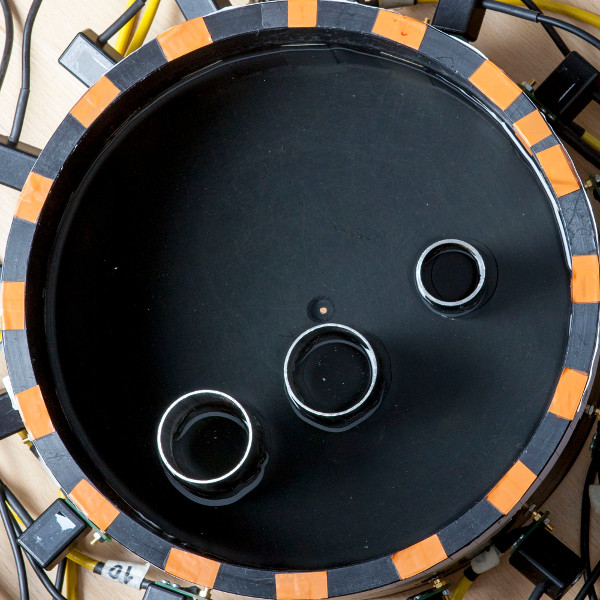} 
  {\bf Case 3.6:} 3 metallic inclusions (circular, hollow)
    \end{minipage}
\caption{Photographs and descriptions of the targets, Cases 3.1 -- 3.6.
\label{fig.photos3}
}
\end{figure}

\begin{figure}[!]
    \centering
    \begin{minipage}{0.25\textwidth}
        \centering
        \includegraphics[width=3cm]{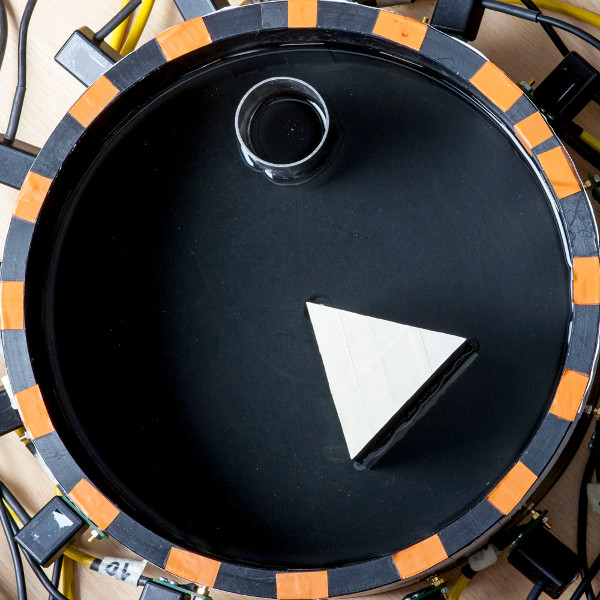} 
  {\bf Case 4.1:} metallic (circular, hollow) and plastic (triangular) inclusion
\vspace{5mm}
    \end{minipage}
    \begin{minipage}{0.25\textwidth}
        \centering
        \includegraphics[width=3cm]{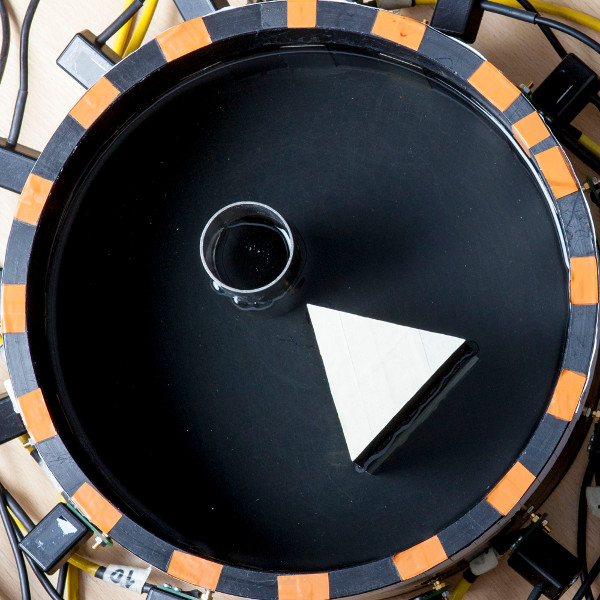} 
  {\bf Case 4.2:} metallic (circular, hollow) and plastic (triangular) inclusion
\vspace{5mm}
    \end{minipage}
    \begin{minipage}{0.25\textwidth}
        \centering
        \includegraphics[width=3cm]{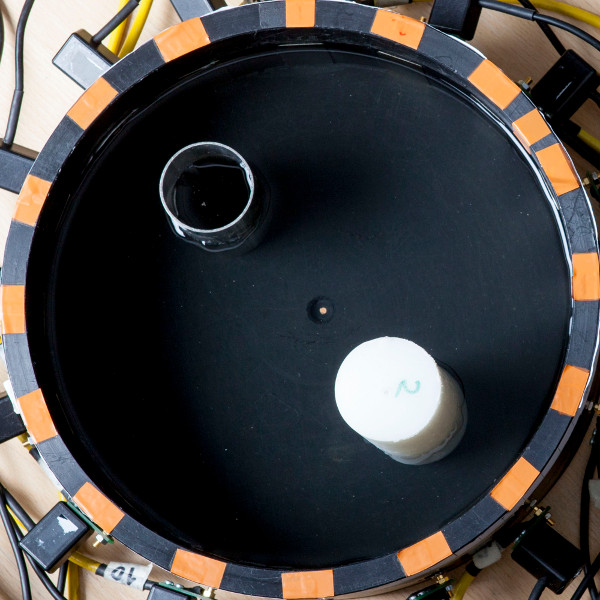} 
  {\bf Case 4.3:} metallic (circular, hollow) and plastic (circular) inclusion
\vspace{5mm}
    \end{minipage}
    \begin{minipage}{0.25\textwidth}
        \centering
        \includegraphics[width=3cm]{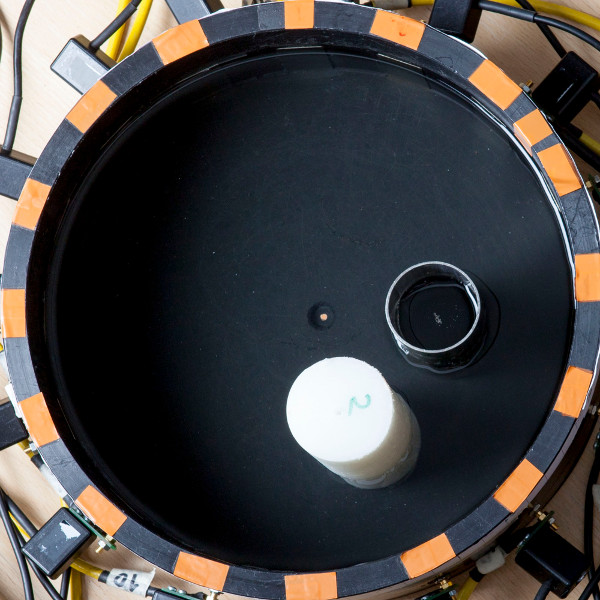} 
  {\bf Case 4.4:} metallic (circular, hollow) and plastic (circular) inclusion
    \end{minipage}
    \begin{minipage}{0.25\textwidth}
        \centering
        \includegraphics[width=3cm]{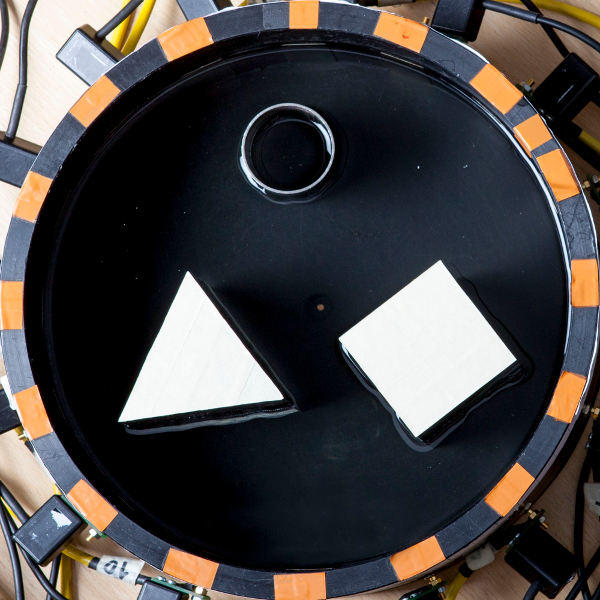} 
  {\bf Case 5.1:} metallic (circular, hollow) and 2 plastic (triangular and rectangular) inclusions
    \end{minipage}
    \begin{minipage}{0.25\textwidth}
        \centering
        \includegraphics[width=3cm]{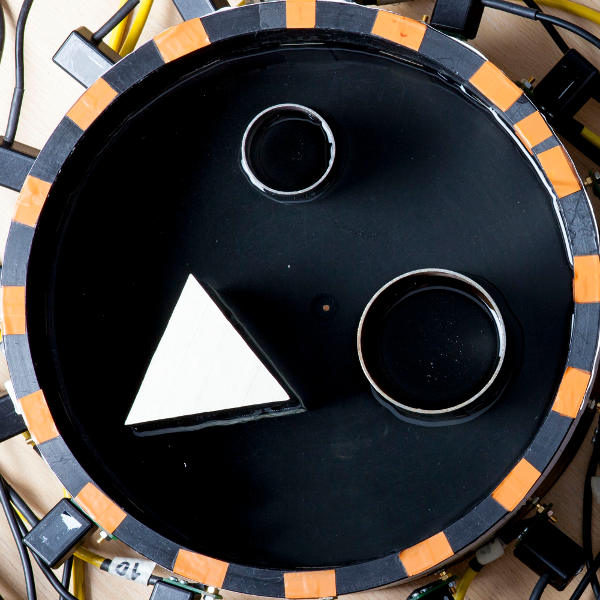} 
  {\bf Case 5.2:} 2 metallic (circular, hollow) and 1 plastic (triangular) inclusion
    \end{minipage}
\caption{Photographs and descriptions of the targets, Cases 4.1 -- 4.4, 5.1 and 5.2.
\label{fig.photos4.5}
}
\end{figure}

\begin{figure}[!]
    \centering
    \begin{minipage}{0.25\textwidth}
        \centering
        \includegraphics[width=3cm]{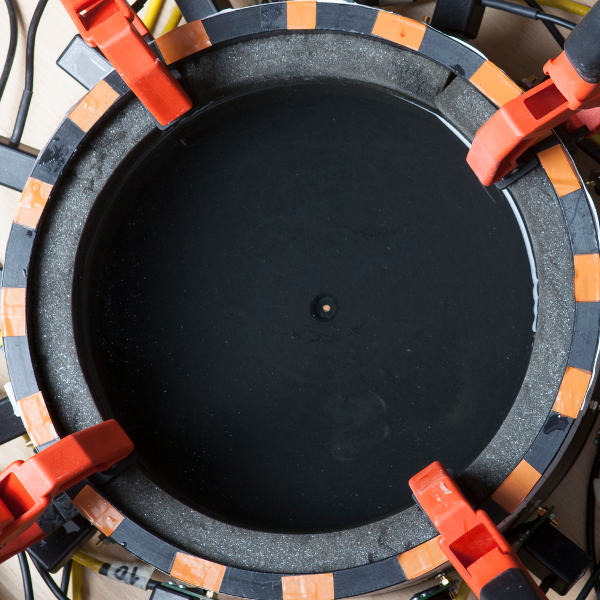} 
  {\bf Case 6.1:} Annular foam layer on the boundary
\vspace{8.5mm}
    \end{minipage}
    \begin{minipage}{0.25\textwidth}
        \centering
        \includegraphics[width=3cm]{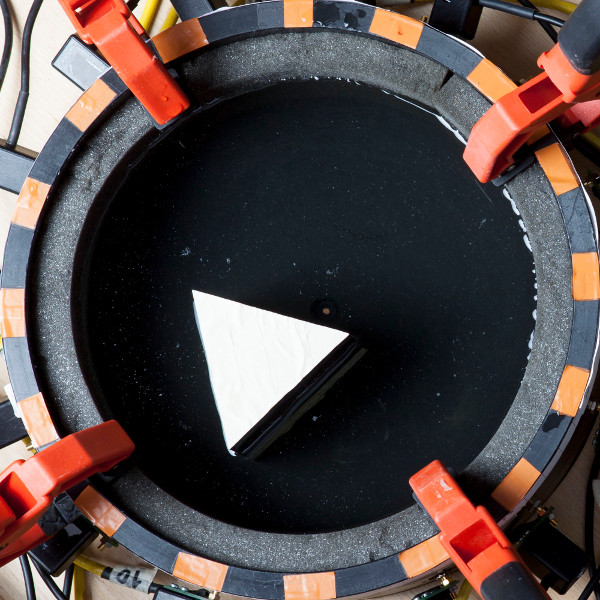} 
  {\bf Case 6.2:} Foam on the boundary. Plastic inclusion (triangular)
\vspace{8.5mm}
    \end{minipage}
    \begin{minipage}{0.25\textwidth}
        \centering
        \includegraphics[width=3cm]{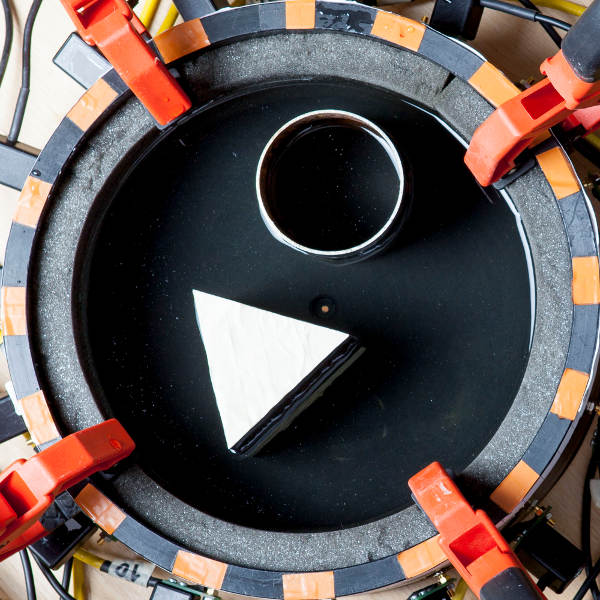} 
  {\bf Case 6.3:} Foam on the boundary. Plastic (triangular) and metallic (circular, hollow) inclusion
    \end{minipage}
    \begin{minipage}{0.25\textwidth}
        \centering
        \includegraphics[width=3cm]{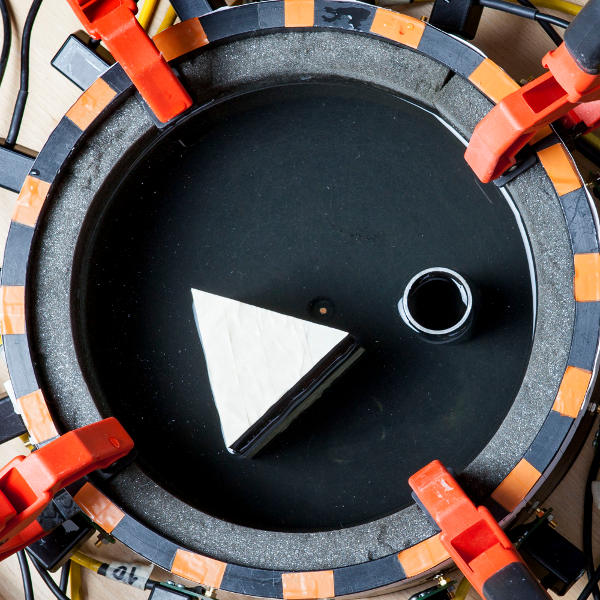} 
  {\bf Case 6.4:} Foam on the boundary. Plastic (triangular) and metallic (circular, hollow) inclusion
%\vspace{5mm}
    \end{minipage}
    \begin{minipage}{0.25\textwidth}
        \centering
        \includegraphics[width=3cm]{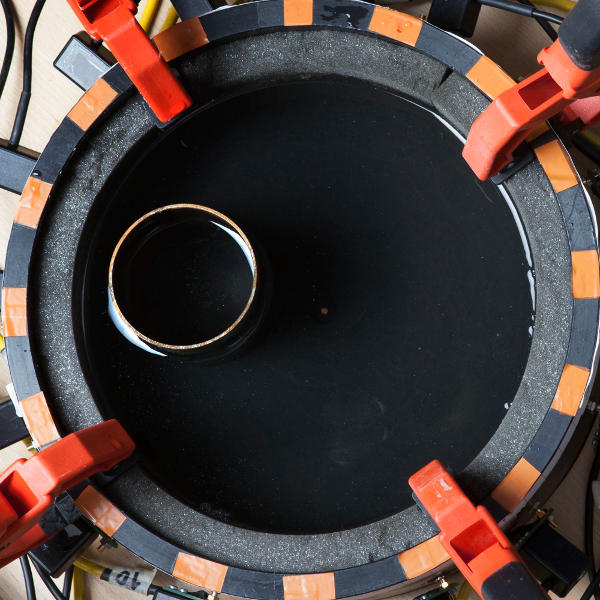} 
  {\bf Case 6.5:} Foam on the boundary. Metallic (circular, hollow) inclusion
\vspace{5mm}
    \end{minipage}
    \begin{minipage}{0.25\textwidth}
        \centering
        \includegraphics[width=3cm]{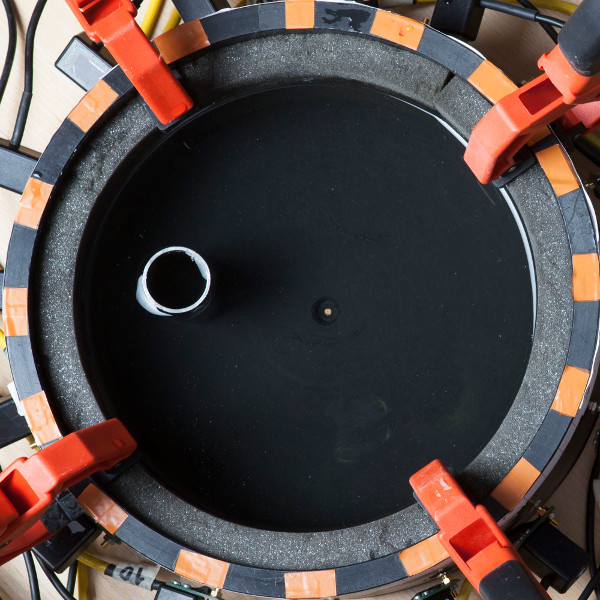} 
  {\bf Case 6.6:} Foam on the boundary. Metallic (circular, hollow) inclusion
\vspace{5mm}
    \end{minipage}
    \begin{minipage}{0.25\textwidth}
        \centering
        \includegraphics[width=3cm]{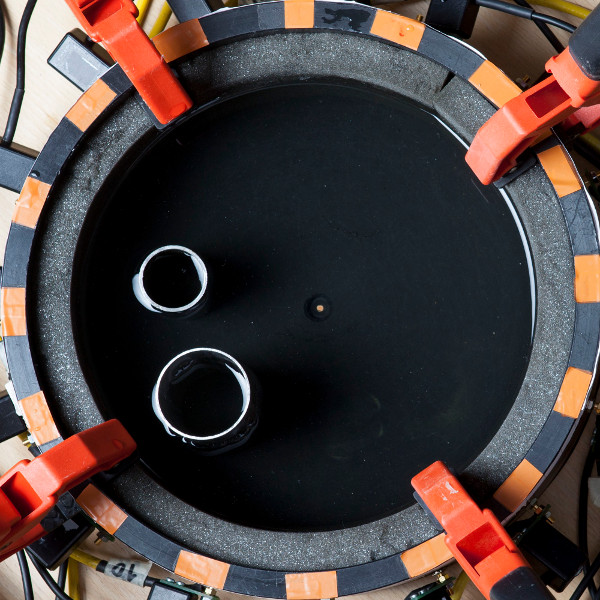} 
  {\bf Case 6.7:}  Foam on the boundary. 2 metallic (circular, hollow) inclusions
    \end{minipage}
    \begin{minipage}{0.25\textwidth}
        \centering
        \includegraphics[width=3cm]{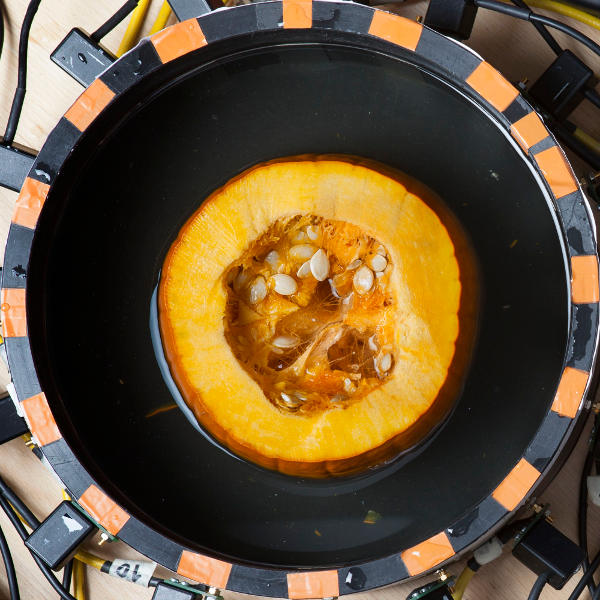} 
  {\bf Case 7.1:} Slice of pumpkin
\vspace{8.5mm}
    \end{minipage}
    \begin{minipage}{0.25\textwidth}
        \centering
        \includegraphics[width=3cm]{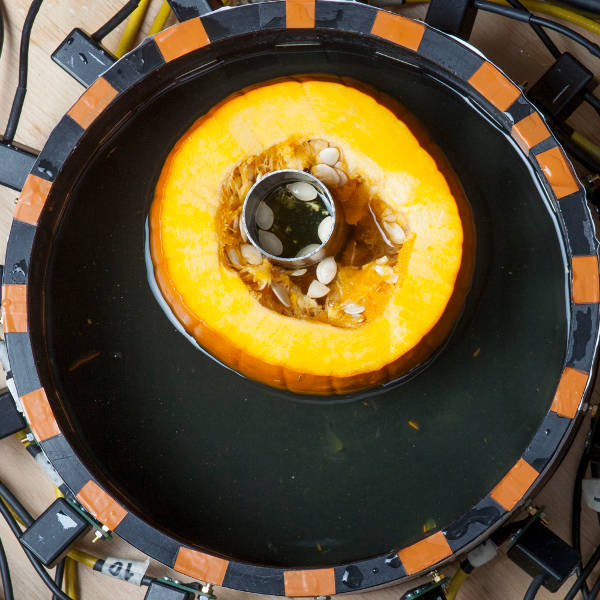} 
  {\bf Case 7.2:} Slice of pumpkin and metallic inclusion (circular, hollow) inside it
    \end{minipage}
\caption{Photographs and descriptions of the targets, Cases 6.1 -- 6.6, 7.1 and 7.2.
\label{fig.photos6.7}
}
\end{figure}

\begin{figure}[!]
    \centering
    \begin{minipage}{0.25\textwidth}
        \centering
        \includegraphics[width=3cm]{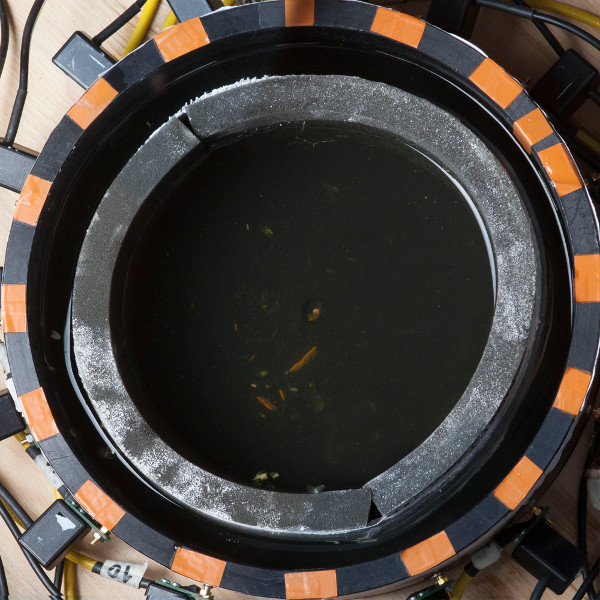} 
  {\bf Case 8.1:} Annular foam near the boundary
\vspace{5mm}
    \end{minipage}
    \begin{minipage}{0.25\textwidth}
        \centering
        \includegraphics[width=3cm]{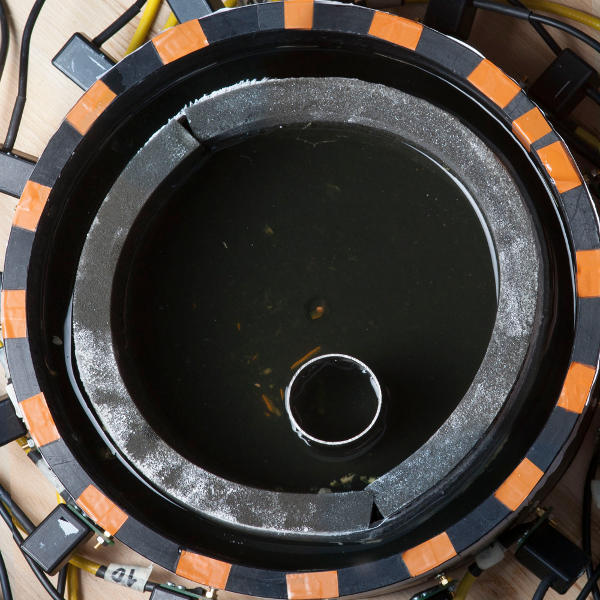} 
  {\bf Case 8.2:} Foam near boundary, metallic (circular, hollow) inclusion
    \end{minipage}
    \begin{minipage}{0.25\textwidth}
        \centering
        \includegraphics[width=3cm]{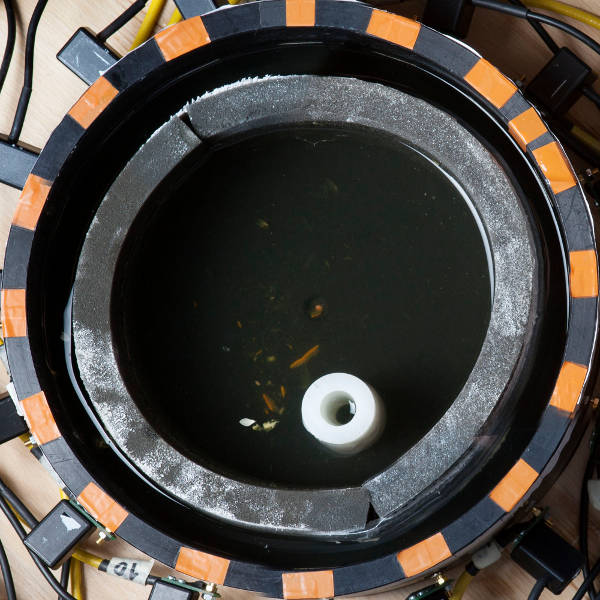} 
  {\bf Case 8.3:} Foam near boundary, plastic (circular, hollow) inclusion
    \end{minipage}
    \begin{minipage}{0.25\textwidth}
        \centering
        \includegraphics[width=3cm]{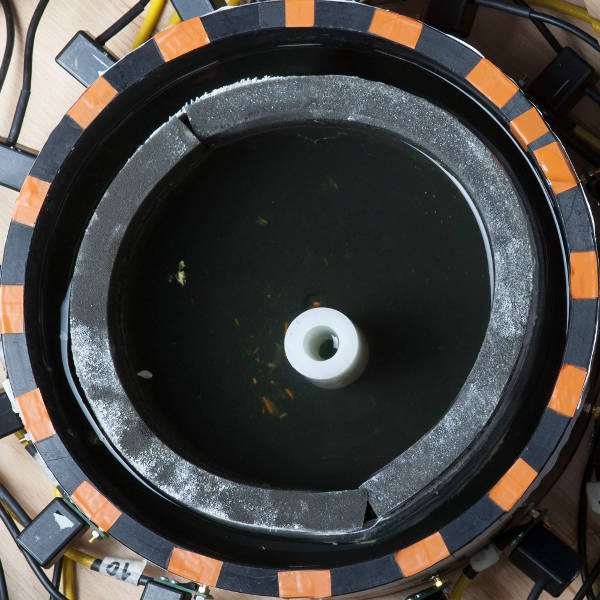} 
  {\bf Case 8.4:} Foam near boundary, plastic (circular, hollow) inclusion
%\vspace{5mm}
    \end{minipage}
    \begin{minipage}{0.25\textwidth}
        \centering
        \includegraphics[width=3cm]{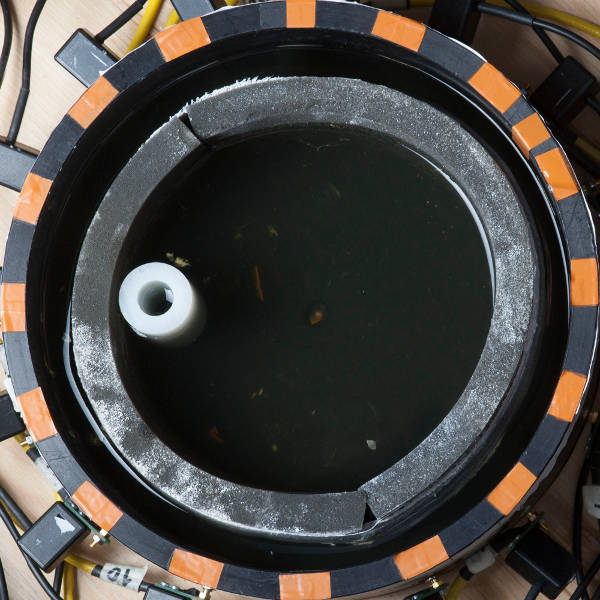} 
  {\bf Case 8.5:} Foam near boundary, plastic (circular, hollow) inclusion
    \end{minipage}
    \begin{minipage}{0.25\textwidth}
        \centering
        \includegraphics[width=3cm]{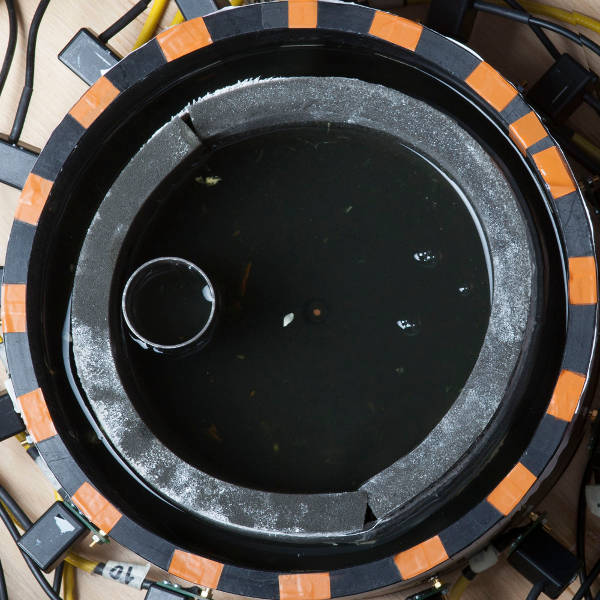} 
  {\bf Case 8.6:} Foam near boundary, metallic (circular, hollow) inclusion
    \end{minipage}
\caption{Photographs and descriptions of the targets, Cases 8.1 -- 8.6.
\label{fig.photos8}
}
\end{figure}

The tank was filled with saline up to the height of 7 cm, i.e., to the top level of the electrodes.
The measured value of the conductivity was $300~{\mu}$S and temperature was $19$ °C.   
Various inclusions were added to the tank in the set of experiments.
%Figure \ref{fig:2a} shows the top view of the tank before adding the inclusions,
%and Figure \ref{fig:2b} corresponds to a case with two non-conductive (plastic) inclusions. 
%%For description of the targets used in the experiments, see Section \ref{Sec:TestCases}.
%
These inclusions are briefly described along their 
photographs in Figures \ref{fig.photos1} -- \ref{fig.photos8}.

%%%%%%%%%%%%%%%%%%%%%
%%%%%%%%%%%%%%%%%%%%%
%%%%%%%%%%%%%%%%%%%%%

%%%%%%%%%%%%%%%%%%%%%
%%%%%%%%%%%%%%%%%%%%%
%%%%%%%%%%%%%%%%%%%%%

\begin{figure}[htp]
	\begin{center}
       \includegraphics[width=14cm]{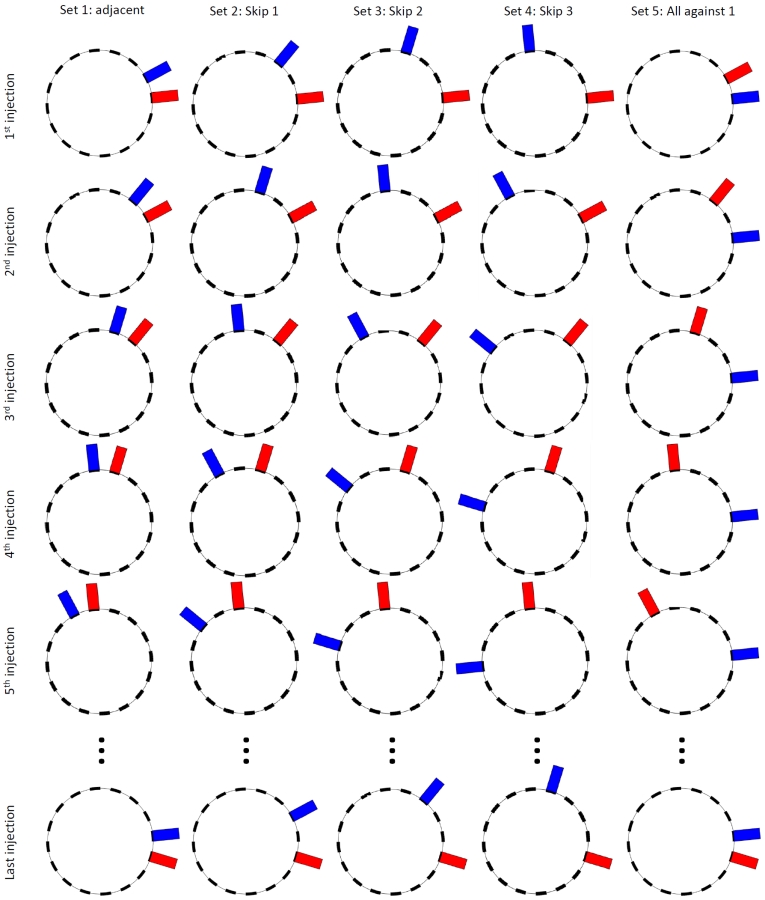}
	\end{center}
	\caption{Current injection patterns used in the experiments. The red and blue bars on the electrodes indicate positive and negative currents through the electrodes, respectively.}
	\label{fig:currents}
\end{figure}

\newpage
For the EIT measurements, a total of $79$ pairwise current injections were used.
The injected currents can be divided into five sets:
\begin{itemize}
\item {\bf Set 1: Adjacent injections.} Injections between electrodes 1-2, 2-3,$\ldots$,15-16, 16-1.
\item {\bf Set 2: Skip 1.} Injections between electrodes 1-3, 2-4,$\ldots$,14-16, 15-1, 16-2.
\item {\bf Set 3: Skip 2.} Injections between electrodes 1-4, 2-5,$\ldots$,13-16, 14-1, $\ldots$, 16-3.
\item {\bf Set 4: Skip 3.} Injections between electrodes 1-5, 2-6,$\ldots$,12-16, 13-1, $\ldots$, 16-4.
\item {\bf Set 5: All against 1.} Injections between electrodes $i$-1, where the $i=2,\ldots,16$.
\end{itemize}
Here, the first electrode $j$ in the electrode pair $j$-$k$ refers to the electrode carrying a positive current and the second electrode $k$ carries the negative current.
The current injections are illustrated in Figure \ref{fig:currents}.

The amplitudes of the currents were $2$ mA and their frequencies were 1 kHz.
Corresponding to each current injection, voltages were measured between all adjacent electrodes: 1-2, 2-3,$\ldots$,15-16, 16-1, resulting in total of 79$\times$16 =1264 measurements. 
%This frame of measurements was repeated 100 times, and saved in a 1264$\times$100 matrix \texttt{Uel} in the case of each target.

%Currents of amplitude $2$mA and frequency $10000$Hz were injected through some specific electrodes follow the Adjacent-Adjacent method ($16$ current projections), Skip 1 ($16$ current projections), Skip 2 ($16$ current projections), Skip 3 ($16$ current projections) and Opposite method ($15$ current projections). This results in $79$ current injections and the corresponding current patterns of the $79$ current injections of an EIT system of $16$ electrodes were stored in the matrix \texttt{I}. Voltage measurements then were transferred to an additional PC connected to the controller module via Ethernet. Voltage measurements were measured at all electrodes, i.e. for 16 current projections of the Adjacent-Adjacent method, one gets $256 \times 1$ voltage measurements. The corresponding voltage measurements of the $79$ current patterns of an EIT system of $16$ electrodes at $100$ different time frames were stored in the matrix \texttt{Uel}. 

%%%%%%%%%%%%%%%%%%%%%%
%%%%%%%%%%%%%%%%%%%%%%

%\section{Test cases}\label{Sec:TestCases}

%In the set of experiments, multiple conductive and resistive inclusions were inserted in to the tank
%These inclusions are briefly described along their 
%photographs in Figures \ref{fig.photos1} -- \ref{fig.photos8}.

%%%%%%%%%%%%%%%%%%%%%%
%%%%%%%%%%%%%%%%%%%%%%

\section{Contents of the data set}
\label{sec.Data}

To use Open 2D EIT data, download and extract the following compressed folders:
\begin{itemize}
\item {\bf Measurements:} \url{fips.fi/EIT_data/data_mat_files.zip}
\item {\bf Photographs:} \url{fips.fi/EIT_data/target_photos.zip}
\end{itemize}
These folders contain, respectively, data files and  photographs  named as 
\begin{itemize}
\item\texttt{datamat$\_$X$\_$Y.mat}
\item \texttt{fantom$\_$X$\_$Y.jpg}
\end{itemize}
where the indices $X$ and $Y$ refer to numbers of the experiments
(cf. Figures \ref{fig.photos1} -- \ref{fig.photos8}).
For example, \texttt{datamat$\_$1$\_$2.mat} stores EIT measurements of Case $1.2$ and \texttt{fantom$\_$1$\_$2.jpg} is a photo of this experiment. 

%The Open 2D EIT data consists of the following files corresponding to each experiment:
%\begin{itemize}
%\item {\bf Measurements:} \texttt{datamat$\_$X$\_$Y.mat} (available in \url{fips.fi/EIT_data/data_mat_files.zip})
%\item {\bf Photograph:} \texttt{fantom$\_$X$\_$Y.jpg} (available in \url{fips.fi/EIT_data/target_photos.zip})
%\end{itemize}
%where X$\_$Y denotes the name of the experiment. For example, \texttt{Data$\_$1$\_$2.mat} stores the current and voltage measurements of the experiment $1.2$ and \texttt{fantom$\_$1$\_$2.jpg} is a photo of this phantom experiment.  Always, \texttt{Data$\_$X$\_$0.mat} denotes the measurements of the reference background.
\newpage

Each data file 
\texttt{datamat$\_$X$\_$Y.mat} contains the following matrices:
\begin{itemize}
\item \texttt{Uel} $ \in\mathbb{R}^{16\times79}$: voltage measurements
(each column consist of 16 adjacent voltage measurements corresponding to one current injection).
\item  \texttt{CurrentPattern} $ \in\mathbb{R}^{16 \times 79}$: current patterns
(each column defines the currents on 16 electrodes in one current injection).
\item  \texttt{MeasPattern} $ \in\mathbb{R}^{16 \times 16}$: measurement patterns
(each column defines the measuring electrodes in the corresponding voltage measurement).
\end{itemize}
A brief Matlab code for reading the data files is provided in link
\url{http://fips.fi/EIT_data/LoadData.m}.
For examples of selecting only parts of data
(e.g., measurements corresponding to any of the current injection sets in Figure \ref{fig:currents})
we refer to comments in the m-file.

Finally, we would like to remind the readers to kindly cite this document whenever using any part of the Open 2D EIT data in a publication.

%Details on the EIT measurements are described in Section \ref{Sec:measurements} above.

%For example, to access the voltage measurements of the Skip 1 method at the $9$th time frame, one can write as follows in MATLAB:
%
%\begin{center}
%\begin{verbatim}
%U = reshape(Uel(:,9),Nmeas,Ninj);
%U = U(:,17:32);
%\end{verbatim}
%\end{center}

%%%%%%%%%%%%%%%%%%%%%%
%%%%%%%%%%%%%%%%%%%%%%

\section{Example reconstructions}
\label{sec.ExampleReconstructions}

\begin{figure}[htp]
	\begin{center}
       \includegraphics[width=14cm]{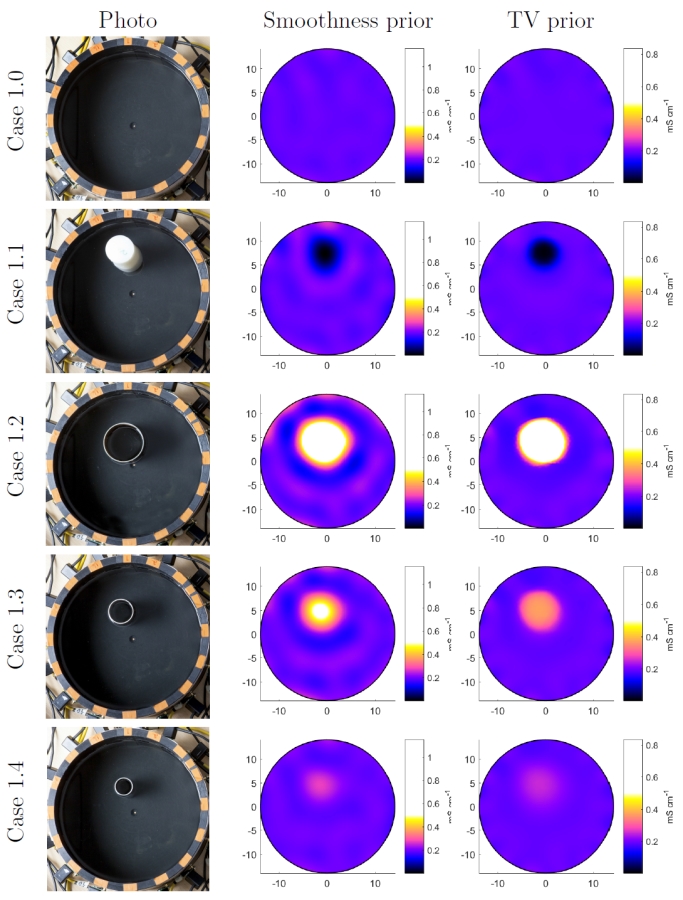}
	\end{center}
	\caption{Open 2D EIT: Cases 1.0 -- 1.4. The left column shows the photographs of the targets, and the middle and right column represent reconstructions with a smoothness prior and a TV prior, respectively.}
	\label{figCase1}
\end{figure}

\begin{figure}[htp]
	\begin{center}
       \includegraphics[width=14cm]{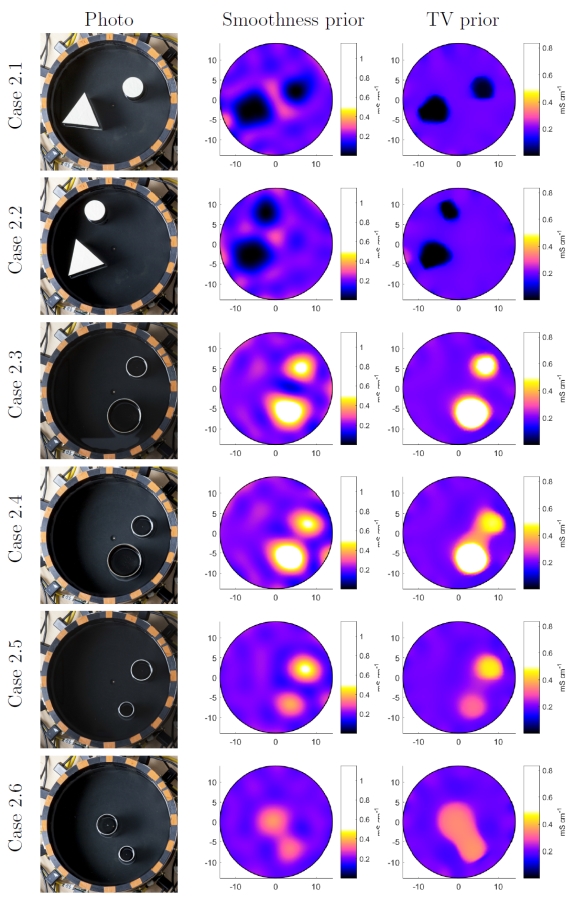}
	\end{center}
	\caption{Open 2D EIT: Cases 2.1 -- 2.6. The left column shows the photographs of the targets, and the middle and right column represent reconstructions with a smoothness prior and a TV prior, respectively.}
	\label{figCase2}
\end{figure}

\begin{figure}[htp]
	\begin{center}
       \includegraphics[width=14cm]{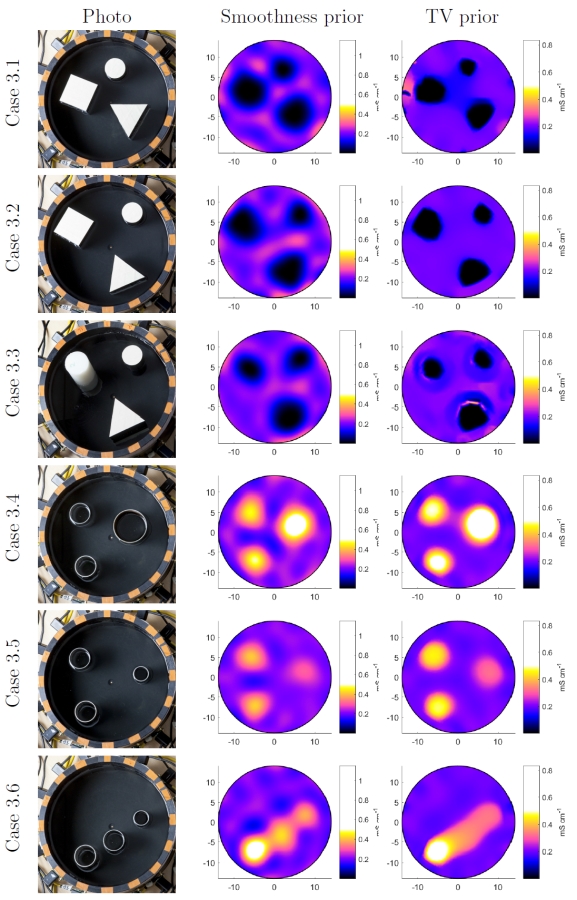}
	\end{center}
	\caption{Open 2D EIT: Cases 3.1 -- 3.6. The left column shows the photographs of the targets, and the middle and right column represent reconstructions with a smoothness prior and a TV prior, respectively.}
	\label{figCase3}
\end{figure}

\begin{figure}[htp]
	\begin{center}
       \includegraphics[width=14cm]{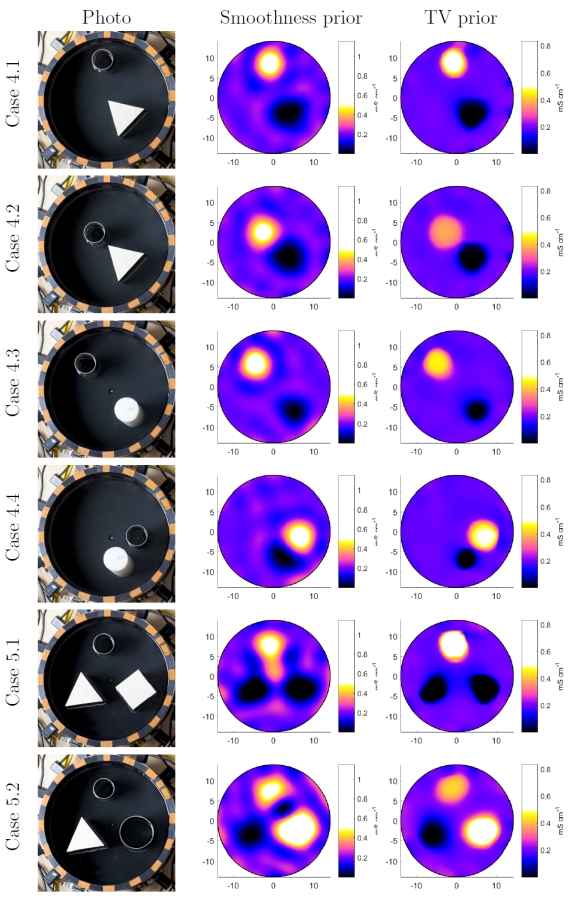}
	\end{center}
	\caption{Open 2D EIT: Cases 4.1 -- 4.4., 5.1 and 5.2. The left column shows the photographs of the targets, and the middle and right column represent reconstructions with a smoothness prior and a TV prior, respectively.}
	\label{figCase4and5}
\end{figure}

\begin{figure}[htp]
	\begin{center}
       \includegraphics[width=12.8cm]{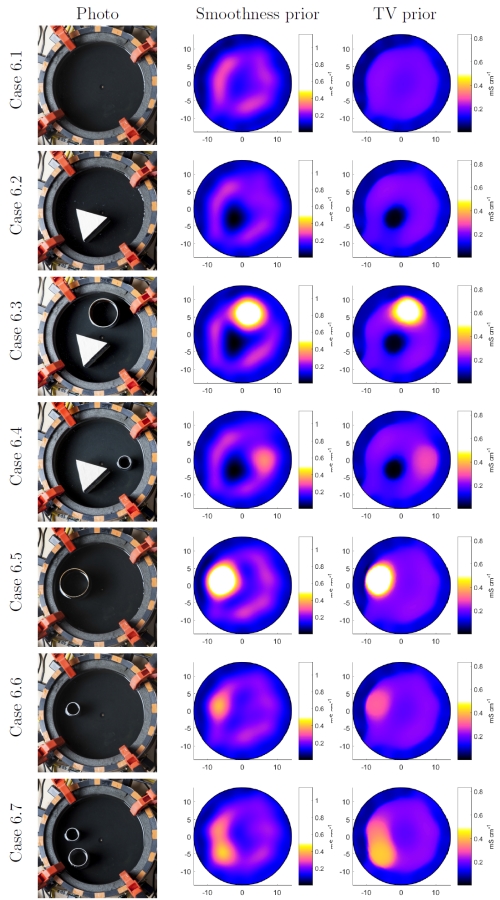}
	\end{center}
	\caption{Open 2D EIT: Cases 6.1 -- 6.7. The left column shows the photographs of the targets, and the middle and right column represent reconstructions with a smoothness prior and a TV prior, respectively.}
	\label{figCase6}
\end{figure}

\begin{figure}[htp]
	\begin{center}
       \includegraphics[width=13cm]{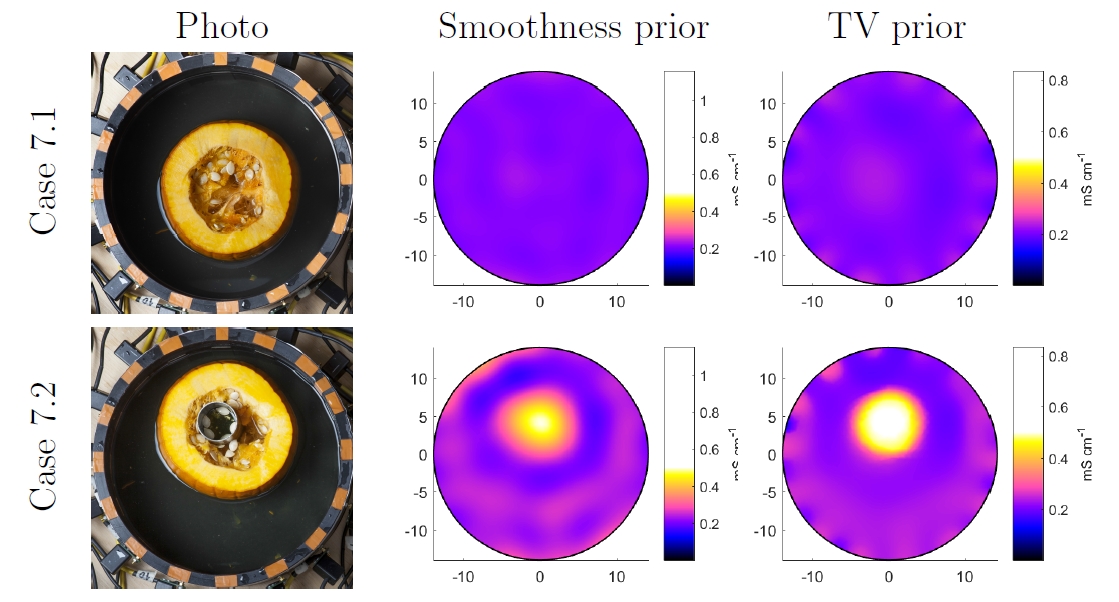}
	\end{center}
	\caption{Open 2D EIT: Cases 7.1 and 7.2. The left column shows the photographs of the targets, and the middle and right column represent reconstructions with a smoothness prior and a TV prior, respectively.}
	\label{figCase7}
\end{figure}

\begin{figure}[htp]
	\begin{center}
       \includegraphics[width=13cm]{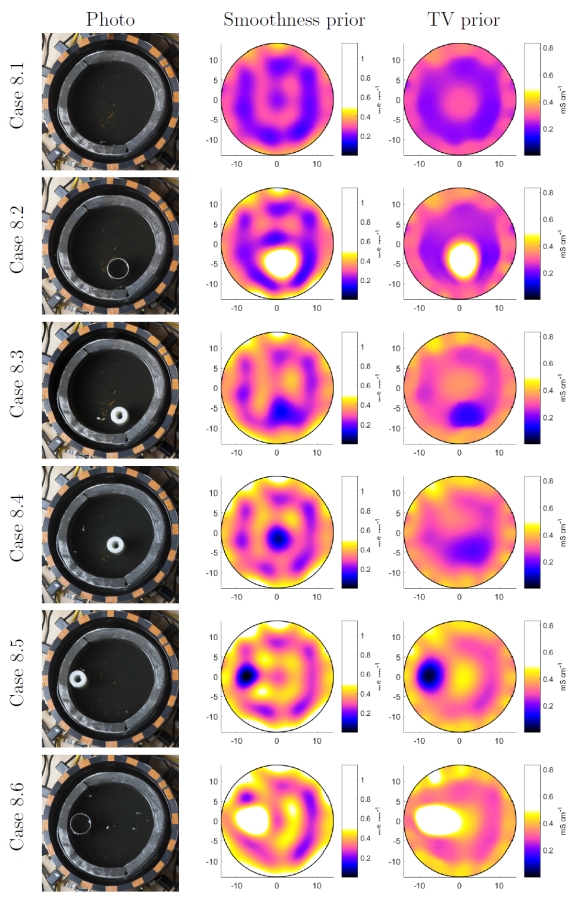}
	\end{center}
	\caption{Open 2D EIT: Cases 8.1 -- 8.6. The left column shows the photographs of the targets, and the middle and right column represent reconstructions with a smoothness prior and a TV prior, respectively.}
	\label{figCase8}
\end{figure}

%The photographs of all targets used in the experiments are shown in Figures \ref{figCase1} -- \ref{figCase8}.
%Each photograph is accompanied with two example reconstructions: 
To illustrate the use of the data, we computed EIT reconstructions
of all targets.
In Figures \ref{figCase1} -- \ref{figCase8},
the photograph of each target is accompanied with two example reconstructions: 
one using a smoothness promoting prior/regularization \cite{lipponen2013}
and one using isotropic TV prior \cite{gonzalez2016}.
All these reconstructions were computed using 
{\em absolute imaging}, i.e., no reference data from the case of homogeneous target was used
in the reconstructions of inhomogeneous conductivity distributions.
For modeling the measurements, 
a three-dimensional (3D) finite element based forward model was used \cite{vauhkonen1999};
however, as all targets were known to be (at least approximately) constant along the 
vertical direction, the conductivity distribution was parametrized in 2D.
Further, as contact impedances between the electrodes and the saline were unknown,
they were estimated simultaneously with the conductivity in each reconstruction
\cite{vilhunen2002}.
We also note that all reconstructions corresponding to each reconstruction method
were computed using the same choices of parameters in the models,
and in Figures \ref{figCase1} -- \ref{figCase8}, all images corresponding to each reconstruction method
are represented in mutually equal color scale.
As the purpose of the reconstructions is only to demonstrate the feasibility of
the measurement data, we omit the details of the reconstruction methods here.
For details, we refer to the papers cited above.

All inclusions in the phantoms are tracked with both reconstruction methods.
In the locations of electrically resistive inclusions,
the reconstructed conductivity is close to zero (indicated by black color),
and in the locations of the highly conductive objects, the 
reconstructed conductivity is higher than the background conductivity (colors ranging from magenta to white).
We note here that the top of the color scale is specifically chosen (wide range of conductivities marked with white) for the purpose of illustrating all reconstructions -- featuring both positive and negative changes from the background conductivity -- in the same color scale. The sensitivity of EIT to the magnitude of the positive conductivity change from the background decreases after certain limit, and consequently in Figures \ref{figCase1} -- \ref{figCase8}, the reconstructed values in the "white range" ($~0.5,\ldots,1.2$ mS/cm) possess high uncertainty.

\bibliographystyle{abbrv}
\bibliography{mybib}

\Addresses
\end{document}